\documentclass[a4paper,11pt,preprintnumbers]{article}
\pdfoutput=1 

\usepackage{fullpage}
\usepackage[T1]{fontenc} 

\usepackage{amssymb}

\begin{document} 
	
\begin{titlepage}
	
	\thispagestyle{empty}
	
	\begin{flushright}
		
		\hfill{CERN-PH-TH-2015-222 \\[1mm] CPHT-RR037.0915 \\[1mm] DFPD-2015/TH/21} 
	\end{flushright}
	
	\vspace{35pt}
	
	\begin{center}
	    { \LARGE{\bf Minimal scalar-less matter-coupled supergravity}}
		
		\vspace{50pt}
		
		{Gianguido~Dall'Agata$~^{1,2,3}$, Sergio Ferrara$~^{4,5,6}$ and Fabio Zwirner$~^{1,2,4}$}
		
		\vspace{25pt}
		
		{
		$^1${\it  Dipartimento di Fisica e Astronomia ``Galileo Galilei''\\
		Universit\`a di Padova, Via Marzolo 8, I--35131 Padova, Italy}
		
		\vspace{15pt}
		
	    $^2${\it   INFN, Sezione di Padova \\
		Via Marzolo 8, I--35131 Padova, Italy}
		
		\vspace{15pt}
		
		$^3${\it Centre de Physique Th\'eorique, \'Ecole Polytechnique, CNRS, \\
		Universit\'e Paris-Saclay, F--91128 Palaiseau, France}

		\vspace{15pt}
		
		$^4${\it  Theory Unit, Physics Department, CERN\\
		CH--1211 Geneva 23, Switzerland}
		
		\vspace{15pt}
		
		$^5${\it  INFN - Laboratori Nazionali di Frascati \\ 
		Via Enrico Fermi 40,	I--00044 Frascati, Italy}
		}
		
		\vspace{15pt}
		
		$^6${\it Department of Physics and Astronomy, \\
		U.C.L.A., Los Angeles CA 90095--1547, USA}
		
		\vspace{40pt}
		
		{ABSTRACT}
	\end{center}
	
	\vspace{10pt}

We build the minimal supergravity model where the nilpotent chiral goldstino superfield is coupled to a chiral matter superfield, realising a different non-linear representation through a mixed nilpotency constraint.
The model describes the spontaneous breaking of local supersymmetry in the presence of a generically massive Majorana fermion, but in the absence of elementary scalars. The sign and the size of the cosmological constant, the spectrum and the four-fermion interactions are controlled by suitable parameters.

\bigskip


\end{titlepage}

\baselineskip 6 mm

\section{Introduction} 
\label{sec:introduction}

Recently there has been some progress in the embedding of non-linear realisations of $N=1$, $D=4$ supersymmetry in supergravity, spontaneously broken through the super-Higgs mechanism \cite{superhiggs1,superhiggs}.
A simple way to implement non-linear realisations is the use of constrained superfields.
The simplest example is the Volkov--Akulov realisation \cite{Volkov:1973ix}, which can be obtained by imposing a nilpotency constraint $X^2=0$ on the chiral superfield $X$ containing the goldstino \cite{nilpotent}.
Its coupling to supergravity has been recently worked out, both in the superfield formalism \cite{Antoniadis:2014oya, Ferrara:2014kva} and in components \cite{Bergshoeff:2015tra}.
Within supergravity, the couplings of the nilpotent goldstino chiral superfield to unconstrained matter chiral superfields have also been considered in \cite{Dall'Agata:2014oka, Dudas:2015eha, matter}, and some of their properties have been studied.
The most important feature of these models is the replacement of the elementary complex scalar of the goldstino multiplet, the sgoldstino \cite{sgoldstino}, with a goldstino bilinear, and this makes them attractive to build simple semi-realistic models of inflation (for a recent review and references, see e.g. \cite{fersag}).
However, the elementary complex scalars of the unconstrained matter chiral superfields still belong to the physical spectrum.  

In this letter we address, in supergravity, the coupling of the goldstino multiplet  $X \equiv (x, \chi, F^x)$,  obeying the constraint $X^2=0$, to a matter chiral multiplet $Y \equiv (y, \psi, F^y)$, also described by a non-linear representation.
As anticipated in the conclusions of  \cite{Dall'Agata:2014oka}, this can be obtained by a mixed nilpotency constraint $XY=0$, as originally introduced in \cite{Brignole:1997pe} and later extended to a more general context \cite{Komargodski:2009rz} in global supersymmetry.
This constraint removes also the elementary complex scalar of the $Y$ multiplet from the spectrum, replacing it with a suitable fermion bilinear, but keeps the spin-1/2 fermion (and the auxiliary field) of $Y$. 
We can then call this model the minimal scalar-less matter-coupled supergravity.
The superspace action we propose is the coupling to supergravity of a slightly generalised form of the action in \cite{Komargodski:2009rz}.
Namely, we keep the minimal canonical K\"ahler potential,
\begin{equation}
\label{kaehler}
K = |X|^2 + |Y|^2 \, ,
\end{equation}
but we introduce the most general superpotential compatible with the superfield constraints imposed on $X$ and $Y$:
\begin{equation}
\label{supo}
W = W_0 + f \, X + g \, Y + h \, Y^2 \, ,
\end{equation}
where it is not restrictive to assume that $(W_0,f,g)$ are real, with $f \ne 0$, and for simplicity we also take $h$ to be real.

We will derive the main physical properties of this minimal model without writing down the full component action.
After solving the nilpotency constraints, the scalar potential will just be a (cosmological) constant,
\begin{equation}
V_0 \equiv \left. V \right|_{x=y=0} = f^2 + g^2 - 3 W_0^2 \, , 
\end{equation} 
with arbitrary sign or vanishing.
For simplicity, we will consider spontaneous supersymmetry breaking in flat space, $V_0=0$, a requirement that imposes a relation between $W_0$, $f$ and $g$, independently of $h$. 
The parameter $g/f$ will measure the $Y$ fraction in the actual goldstino, $\widetilde{G} \propto f \, \chi + g \, \psi$.
The mass of the physical spin-1/2 state will then depend on $W_0$, $g/f$ and $h$.
We will show that, because of the nilpotency constraints, computing the spin-1/2 fermion masses requires a correction to the standard supergravity formula, and provide a simple formula for such a correction, which can also be applied to non-minimal models with an arbitrary number of constrained chiral superfields.
Both $h$ and $g/f$ will also affect the four-fermion interactions.

	
\section{The minimal model with $X^2 = X Y = 0$} 
\label{sec:the_model}

We couple minimal supergravity to two chiral multiplets, $X$ and $Y$, subject to the nilpotency constraints \cite{Brignole:1997pe,Komargodski:2009rz}
\begin{equation}
\label{constraints}
X^2 = X \, Y = 0 \, .
\end{equation}
These constraints fix the lowest components of $X$ and $Y$ in terms of fermion bilinears \cite{Komargodski:2009rz}:
\begin{eqnarray}
\label{expX}
X & = &  \frac{\chi \, \chi}{2 \, F^x} + \sqrt2 \, \theta \, \chi + \theta^2 \, F^x \, ,
\\
\label{expY}
Y & = & \frac{\psi \, \chi}{F^x} - \frac{\chi \, \chi}{2 \, (F^x)^2} \, F^y + \sqrt2 \, \theta \,  \psi + \theta^2 \, F^y \, .
\end{eqnarray}
Notice that, since the constraints in (\ref{constraints}) and their solutions in (\ref{expX}) and (\ref{expY}) are algebraic, there is no problem to implement them in local supersymmetry.
Equation (\ref{constraints}) corresponds, in the superspace action described in \cite{Ferrara:2014kva}, to an $F$-term with two Lagrange multipliers $\Lambda_{1}$ and $\Lambda_2$, multiplying the two corresponding monomials $X^2$ and $XY$.
Notice also that the two solutions in (\ref{expX}) and (\ref{expY}) are singular when the auxiliary field $F^x$ of the goldstino multiplet $X$ vanishes, but regular in the auxiliary field $F^y$ of the matter multiplet $Y$, which can vanish or not depending on the model.
For our minimal model defined by (\ref{kaehler}) and (\ref{supo}), we will see that $F^x |_{x=y=0}= -f$ and $F^y |_{x=y=0}= -g$.
In addition, as the reader can easily check, the solution (\ref{expY}) implies the nilpotency condition $Y^3 = 0$ for the matter superfield $Y$.

Since the scalar components of the two superfields are replaced by fermion bilinears, there is a trick to efficiently compute the two leading terms of the expansion of the full supergravity action in fermion bilinears, i.e. the scalar potential and the fermion mass terms.
We can just take the standard expression of the supergravity Lagrangian and replace the composite scalars $x$ and $y$ with the corresponding fermion bilinears (or set them to zero if we are only interested in the bosonic action), disregarding the four-fermion and higher-order terms generated in the process. 

As a first step, we choose the canonical K\"ahler potential (\ref{kaehler}), with the most general superpotential (\ref{supo}) compatible with the constraints (\ref{constraints}).
The relevant non-zero bosonic quantities are
\begin{equation}\label{DW}
D_X W|_{x=y=0} = f \, , 
\quad 
D_Y W|_{x=y=0} = g \, , 
\quad 
W|_{x=y=0} = W_0 \, , 
\quad 
D_Y D_Y W|_{x=y=0} = 2 h \, ,
\end{equation}
\begin{equation}
\partial_X V|_{x=y=0} = - 2 \, f \, W_0 \, , 
\qquad 
\partial_Y V|_{x=y=0} = 2 \, g \, (h-W_0) \, ,
\end{equation}
where $D_i W = W_i + W \, K_i$ ($i=X,Y$).
This means that there is a non-trivial constant scalar potential in the theory,
\begin{equation}\label{potential}
V_0  = \left[ e^K \left( |D_X W|^2 +  |D_Y W|^2  - 3 \, |W|^2 \right) \right]_{x=y=0} 
= f^2 + g^2 - 3 \, W_0^2 \, ,
\end{equation}
which can be positive, negative or zero, according to the choices of the parameters $f,g,W_0$.

For the time being we fix $W_0$ to obtain a Minkowski vacuum, $V_0 = 0$.
The spectrum of the theory contains only the graviton, the massive gravitino (which absorbs the would-be goldstino) and another physical spin-1/2 Majorana fermion.
The gravitino mass is
\begin{equation}
\label{mgrav}
m_{3/2} = W_0 = \sqrt{\frac{f^2 + g^2}{3}} \, .
\end{equation}
In the spin-1/2 sector, after removing the mixing between goldstino and gravitino with some standard gauge choice, the actual goldstino $\widetilde{G}$ should have vanishing mass, while the orthogonal combination of $\chi$ and $\psi$ may be massive.
It is interesting to notice that, if we naively used the standard spin-1/2 fermion mass matrix of supergravity after projecting out the goldstino, namely
\begin{equation}
M_{ij} = e^{K/2}\, \left[D_i D_j W -\frac23 \frac{D_i W D_j W}{W}\right]_{x=y=0} \, ,
\qquad
(i=X,Y) \, , 
\end{equation}
we would get an incorrect result for the fermion mass spectrum.
In fact the matrix
\begin{equation}
M = \left(\begin{array}{cc}
-\frac23 \frac{f^2}{W_0} & -\frac23 \frac{fg}{W_0} \\[2mm]
-\frac23 \frac{fg}{W_0} & 2h -\frac23 \frac{g^2}{W_0}
\end{array}\right)
\end{equation}
would not exhibit the vanishing eigenvalue corresponding to the goldstino eigenvector.
To compute the correct mass matrix, we should take into account those additional terms that appear in the Lagrangian because the scalar components $x$ and $y$ of the $X$ and $Y$ superfields have been replaced by fermion bilinears.
Each linear term in the scalar potential contributes with a coefficient that depends on the expansions (\ref{expX}) and (\ref{expY}), so that the correct mass formula is 
\begin{equation}
M_{ij} = e^{K/2}\, \left[D_i D_j W -\frac23 \frac{D_i W D_j W}{W}\right]_{s=0} + 2\, \partial_{k} V|_{s^k = 0} 
\, \frac{\partial s^k}{\partial (\chi^i \chi^j)},
\end{equation}
where $s^k = S^k|_{\theta =0}$.
In our minimal model, $S^1=X$ and $S^2=Y$, but the above formula can be generalised to an arbitrary number of constrained chiral multiplets, whose scalar components are replaced by fermion bilinears.
For our model, the last term contributes to the final masses with the following correction 
\begin{equation}
\delta M = \left(
\begin{array}{cc}
2 \, \partial_X V|_{x=y=0} \, \partial_{\chi \chi} \, X + 2 \, \partial_Y V|_{x=y=0} \, \partial_{\chi \chi} Y & \quad \partial_Y V|_{x=y=0} \, \partial_{\chi \psi} Y \\[2mm]
\partial_Y V|_{x=y=0} \, \partial_{\chi \psi} Y & 0
\end{array}
\right) \, ,
\end{equation}
which gives
\begin{equation}
\delta M = \left(
\begin{array}{cc}
2 W_0 - 2 \frac{g^2}{f^2}(W_0-h) & \frac{2 g }{f}(W_0 - h) \\[2mm]
 \frac{2 g }{f}(W_0 - h)& 0
\end{array}
\right) \, .
\end{equation}
Making use of the relations in (\ref{mgrav}), the corrected mass matrix has a vanishing eigenvalue for the goldstino eigenvector and a non-trivial eigenvalue for the orthogonal eigenvector $\widetilde{F} \propto g \, \chi - f \, \psi$: 
\begin{equation}
\label{spinmass}
	m_{1/2}  = 2 \, h \, \left( 1+ \frac{g^2}{f^2} \right) - 2 \, \frac{g^2}{f^2} \, m_{3/2} \, .
\end{equation}
Notice that $m_{1/2}$ can vanish also for non-vanishing values of $g$ and $h$, because the two terms in (\ref{spinmass}) can have opposite sign.

Following the same line of reasoning, we could study the four-fermion couplings of the model, which depend on the free superpotential parameters $W_0$, $f/g$ and $h$. However, such a study would require the systematic derivation of the  fermionic action and goes beyond the aim of the present letter. 


\section{A generalization} 
\label{sec:generalization}

While the minimal model has the canonical K\"ahler potential (\ref{kaehler}), we can study the most general model for two superfields $X$ and $Y$ obeying the constraints (\ref{expX}) and (\ref{expY}) by considering:
\begin{equation}\label{genkahler}
	K = |X|^2 + |Y|^2 + a\, (X {\overline Y}^2 + {\overline X} Y^2) + b\, (Y {\overline Y}^2 + {\overline Y} Y^2) + c \, |Y|^4 \, .
\end{equation}
While the technical details of this model change with respect to the minimal one, the physics remains qualitatively the same.

At $x = y = 0$ we have the same K\"ahler-covariant derivatives of the superpotential, potential and gravitino mass as in (\ref{DW}), (\ref{potential}) and (\ref{mgrav}).
Also the K\"ahler metric remains canonical at $x = y=0$.
The derivatives of the potential change, but only in the $Y$ direction:
\begin{equation}
	\partial_X V|_{x=y=0} = - 2 \, f \, W_0 \, , 
	\qquad 
	\partial_Y V|_{x=y=0} = -2 \, g \, (a\,f + b\,g - h + W_0) \, .
\end{equation}
This change is needed to compensate in the mass formula for the new term that appears in the standard spin-1/2 mass formula from the computation of the Christoffel connection, evaluated at $x=y=0$:
\begin{equation}
	\Gamma^x_{yy}|_{x=y=0} = 2 \, a \, ,
	\qquad 
	\Gamma^y_{yy}|_{x=y=0} = 2 \, b \, .
\end{equation}
The mass of the physical fermion becomes:
\begin{equation}
	m_{1/2}  = 2 \, \frac{\left(f^2+g^2\right)(h-a \, f - b \, g)- g^2 \, m_{3/2}}{f^2} \, .
\end{equation}
Note that the $c$ parameter in the K\"ahler potential (\ref{genkahler}) does not affect the mass spectrum, but it will generate four-fermion matter couplings in the original supergravity Lagrangian, as well as other couplings originating from the solution of the constraints.

\newpage

\section{Conclusions and outlook} 
\label{sec:conclusions_and_outlook}

In this letter we considered the minimal scalar-less model of matter-coupled spontaneously broken supergravity, where in addition to the nilpotent goldstino multiplet also a matter chiral multiplet transforms in a non-linear representation.
The model describes the locally supersymmetric interactions of a massive gravitino and a massive spin-1/2 Majorana fermion. 

Along similar lines, we could build another minimal example by considering the non-linear realisation of an Abelian vector multiplet through the constraint $XW_{\alpha}=0$ \cite{Komargodski:2009rz}, coupled to supergravity with an action that is the sum of the Volkov--Akulov action and the Maxwell action.
This model describes a massive gravitino coupled to the Maxwell massless vector field,
The gaugino is a composite of the vector field strength and of the goldstino, thus it vanishes in the unitary gauge. 
It would be interesting to build more general models with mixed nilpotency constraints among matter fields in chiral and vector multiplets and the goldstino chiral multiplet, and to find the full component expression of their Lagrangians, along the lines of \cite{Bergshoeff:2015tra}.

The revival of the study of non-linear representations of supersymmetry coupled to supergravity was originally motivated by their application to inflationary models \cite{AlvarezGaume:2010rt} (for a review of the vast recent literature on the cosmological applications of nilpotent superfields, and an extensive list of references, see again \cite{fersag}).
Models with nilpotency constraints also enable \cite{Ferrara:2014kva} a consistent four-dimensional effective supergravity description of the so-called KKLT uplift \cite{Kachru:2003aw}, associated with brane supersymmetry breaking \cite{branesb} (the inadequacy of standard supergravity for this purpose was shown long ago in section~2 of \cite{vzD}).
General nilpotency constraints in the framework of the superconformal tensor calculus \cite{Freedman:2012zz} were indicated in \cite{Ferrara:2014kva}.
Some of these models are  dual to higher-curvature supergravities with nilpotency constraints on the chiral curvature multiplet \cite{Antoniadis:2014oya, Dudas:2015eha, Hasegawa:2015era}.
However, the study of mixed nilpotency constraints was proposed until now only in the framework of rigid supersymmetry \cite{Brignole:1997pe, Komargodski:2009rz} or in the superspace formulation of the supersymmetric Born-Infeld action \cite{susyBI}.
In this paper we have presented for the first time a minimal supergravity model of this sort, where the spectrum, apart from the graviton, is purely fermionic.
As already mentioned in the conclusions of \cite{Dall'Agata:2014oka}, it would be interesting to see if such mixed nilpotency constraints can also find applications to cosmology.



\bigskip
\section*{Acknowledgments}

\noindent We would like to thank I.~Antoniadis, E.~Dudas, R.~Kallosh, A.~Kehagias, A.~Linde, M.~Porrati and A.~Sagnotti for discussions or collaborations on related issues.
This work is supported in part by the ERC Advanced Grant no.267985 (DaMeSyFla), by the MIUR-PRIN project 2010YJ2NYW, by the MIUR grant RBFR10QS5J (STaFI), by the INFN-CSN4 (I.S. GSS and HEPCube) and by the Padova University Project CPDA119349.

\newpage

\end{document}